\documentclass[sigconf,nonacm]{acmart}

\usepackage{braket}

\AtBeginDocument{%
  }

\begin{document}

\title[Physics-Informed QAOA for RIS]{Quantum Optimization for Electromagnetics: Physics-Informed QAOA for Reconfigurable Intelligent Surfaces}





\author{Marco Pasquale}
\affiliation{%
  \institution{Computational Science and Technology}
  \institution{KTH Royal Institute of Technology}
  \city{Stockholm}
  \country{Sweden}}
\email{marcopas@kth.se}

\author{Erik M. Åsgrim}
\affiliation{%
  \institution{Computational Science and Technology}
  \institution{KTH Royal Institute of Technology}
  \city{Stockholm}
  \country{Sweden}}
\email{erima@kth.se}

\author{Stefano Markidis}
\affiliation{%
  \institution{Computational Science and Technology}
  \institution{KTH Royal Institute of Technology}
  \city{Stockholm}
  \country{Sweden}}
\email{markidis@kth.se}

\author{Oscar Quevedo-Teruel}
\affiliation{%
  \institution{Communication Systems}
  \institution{KTH Royal Institute of Technology}
  \city{Stockholm}
  \country{Sweden}}
\email{oscarqt@kth.se}

\begin{abstract}
Optimizing Reconfigurable Intelligent Surfaces (RIS) is a high-dimensional combinatorial challenge. Current quantum algorithms often simplify this problem by ignoring physical constraints like mutual coupling, which significantly degrades real-world performance. Rather than targeting a fully realistic RIS description, we embed progressively more physics-informed models of mutual coupling into Quadratic Unconstrained Binary Optimization (QUBO) formulations. We evaluate four Ising interaction models ($J_{ij}$) for the Quantum Approximate Optimization Algorithm (QAOA), ranging from idealized phase-only to fully dense physical models. Analyzing a $5 \times 5$ grid, our results expose a critical trade-off between spatial pointing accuracy and quantum hardware feasibility. While complete global coupling maximizes beamforming precision, dense Hamiltonians introduce prohibitive routing overhead and complicate convergence on near-term processors. Ultimately, we demonstrate that while physics-informed quantum optimization is mathematically viable, sparse, distance-penalized models remain a necessary compromise for execution on current noisy intermediate-scale quantum (NISQ) devices.
\end{abstract}




\begin{teaserfigure}
    \centering
    \raisebox{-0.45\height}{\includegraphics[width=0.32\textwidth, clip=true, trim=0.45cm 0.3cm 0.5cm 0.3cm]{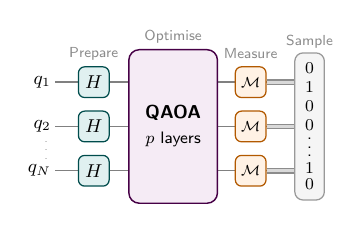}}
    \hfill
    \raisebox{-0.4\height}{%
    \includegraphics[width=0.131\textwidth, keepaspectratio=true, clip=true, trim=6.6cm 4.5cm 5.4cm 5.5cm]{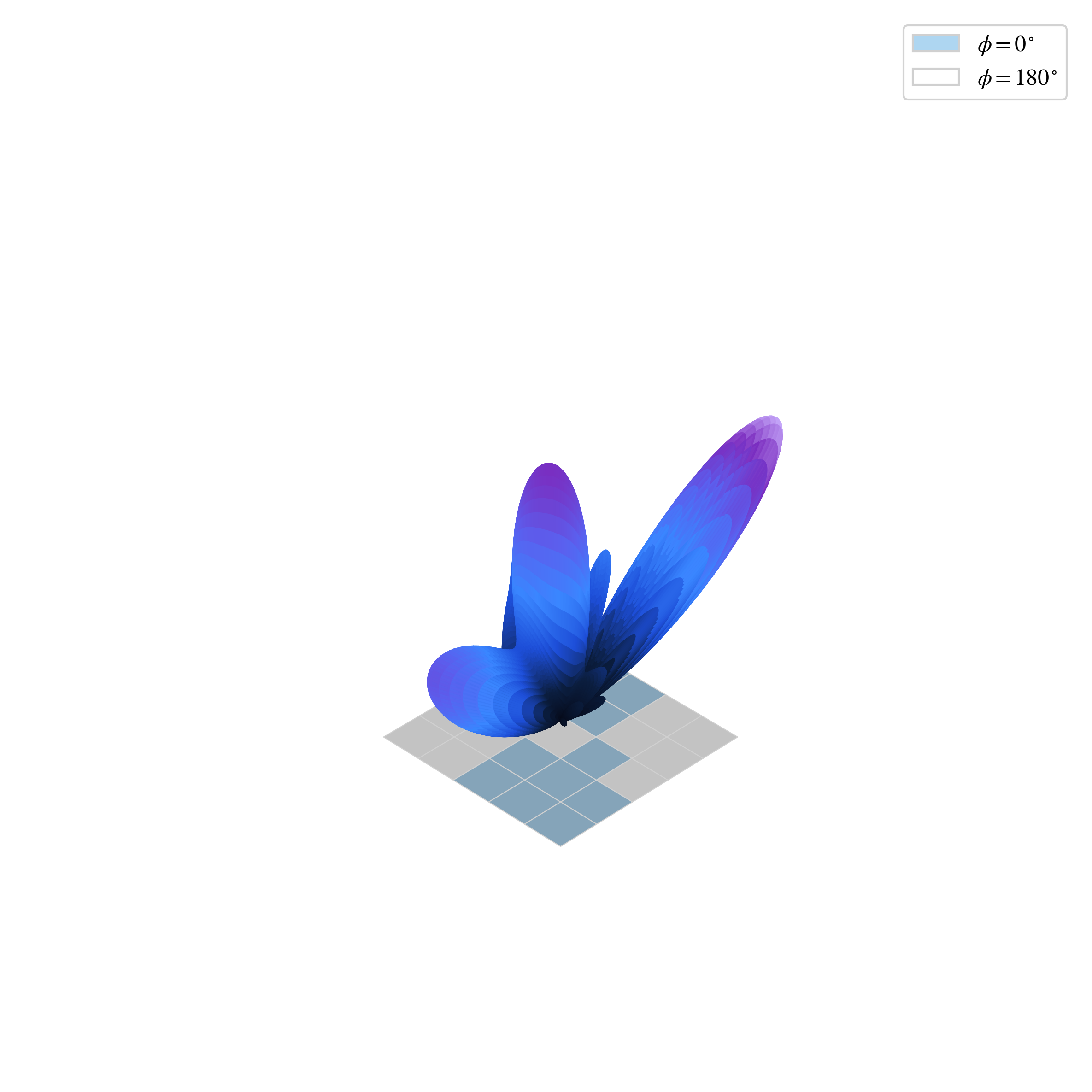}%
    \hspace{-4mm}%
    \includegraphics[width=0.131\textwidth, keepaspectratio=true, clip=true, trim=6.6cm 4.5cm 5.4cm 5.5cm]{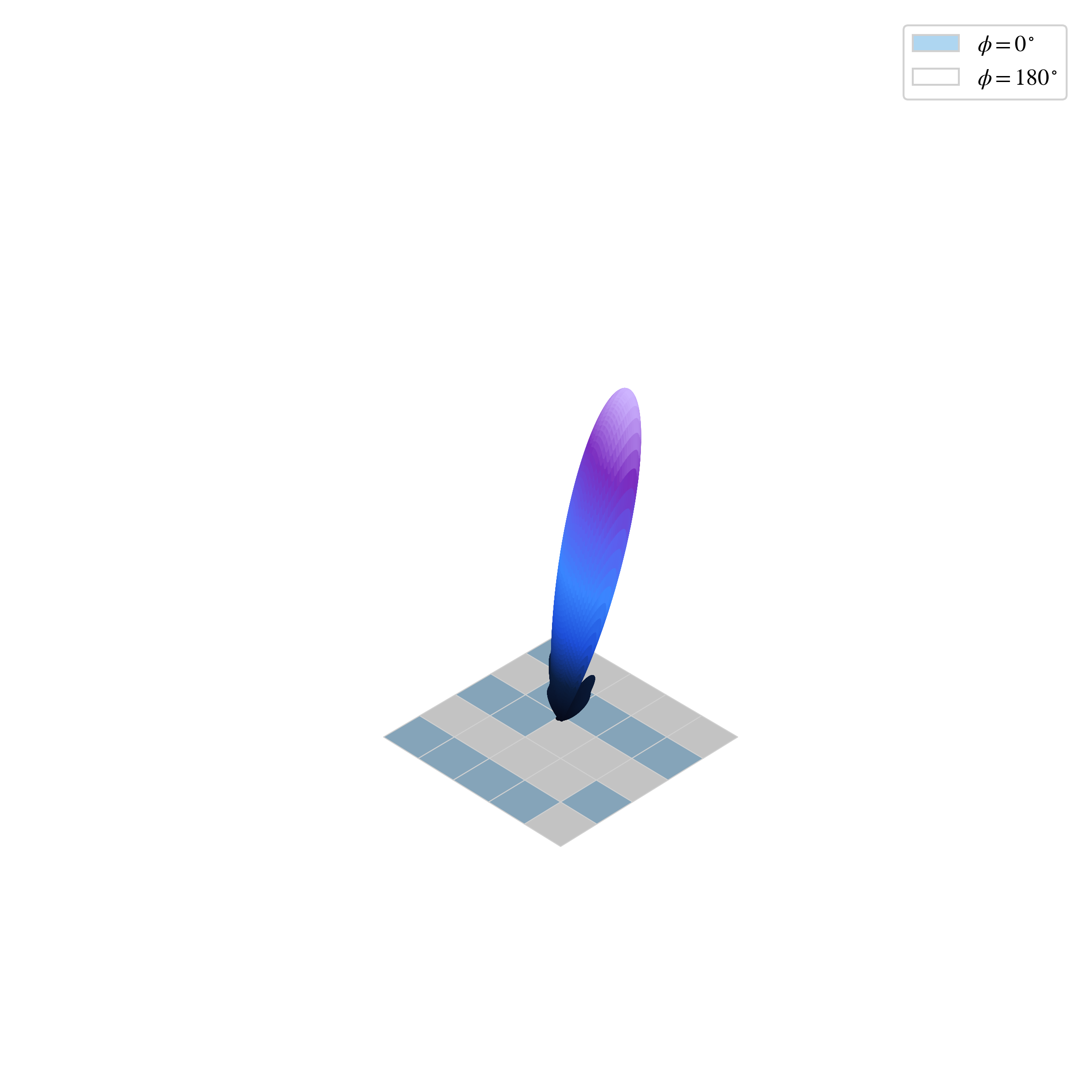}%
    \hspace{-4mm}%
    \includegraphics[width=0.131\textwidth, keepaspectratio=true, clip=true, trim=6.6cm 4.5cm 5.4cm 5.5cm]{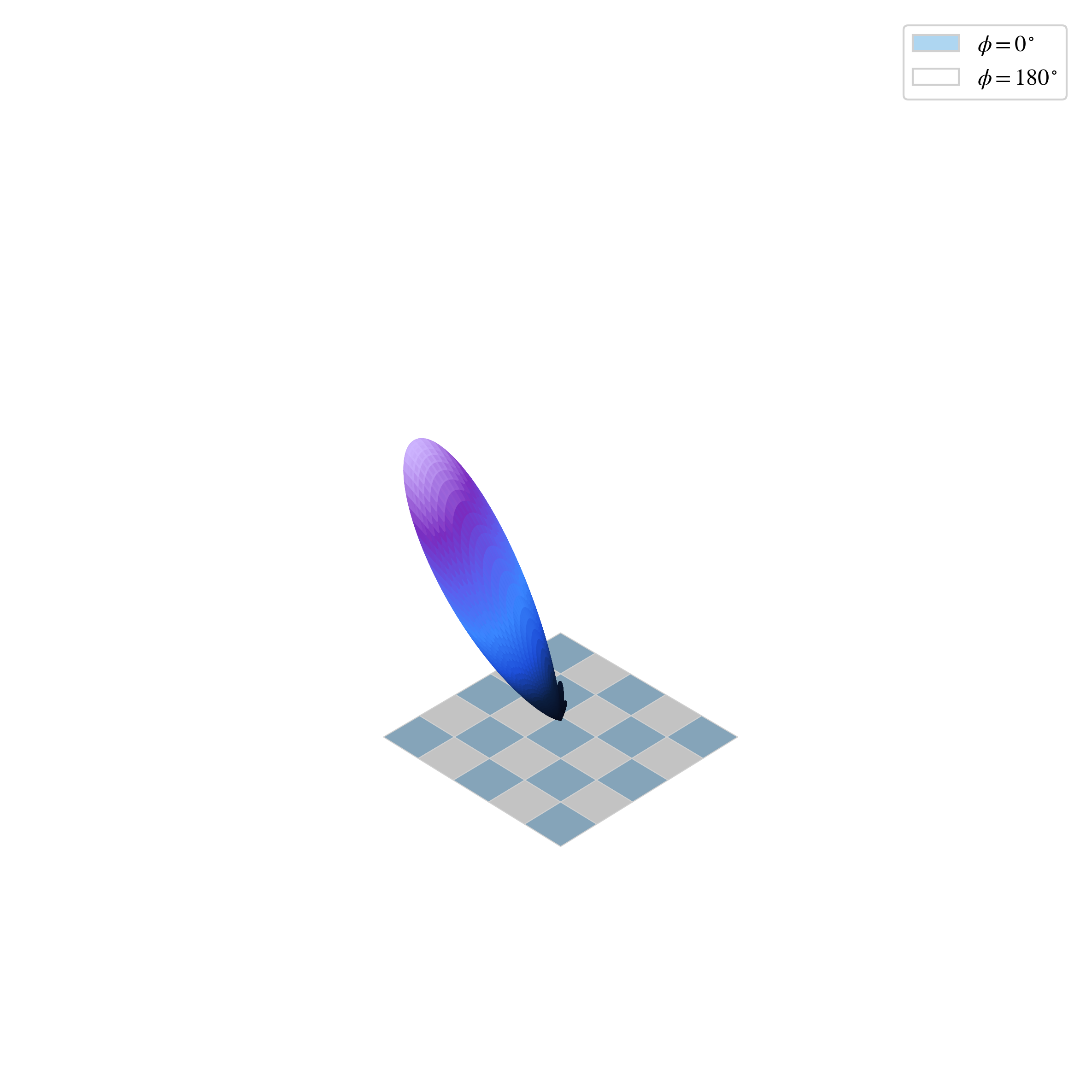}%
    \hspace{-4mm}%
    \includegraphics[width=0.131\textwidth, keepaspectratio=true, clip=true, trim=6.6cm 4.5cm 5.4cm 5.5cm]{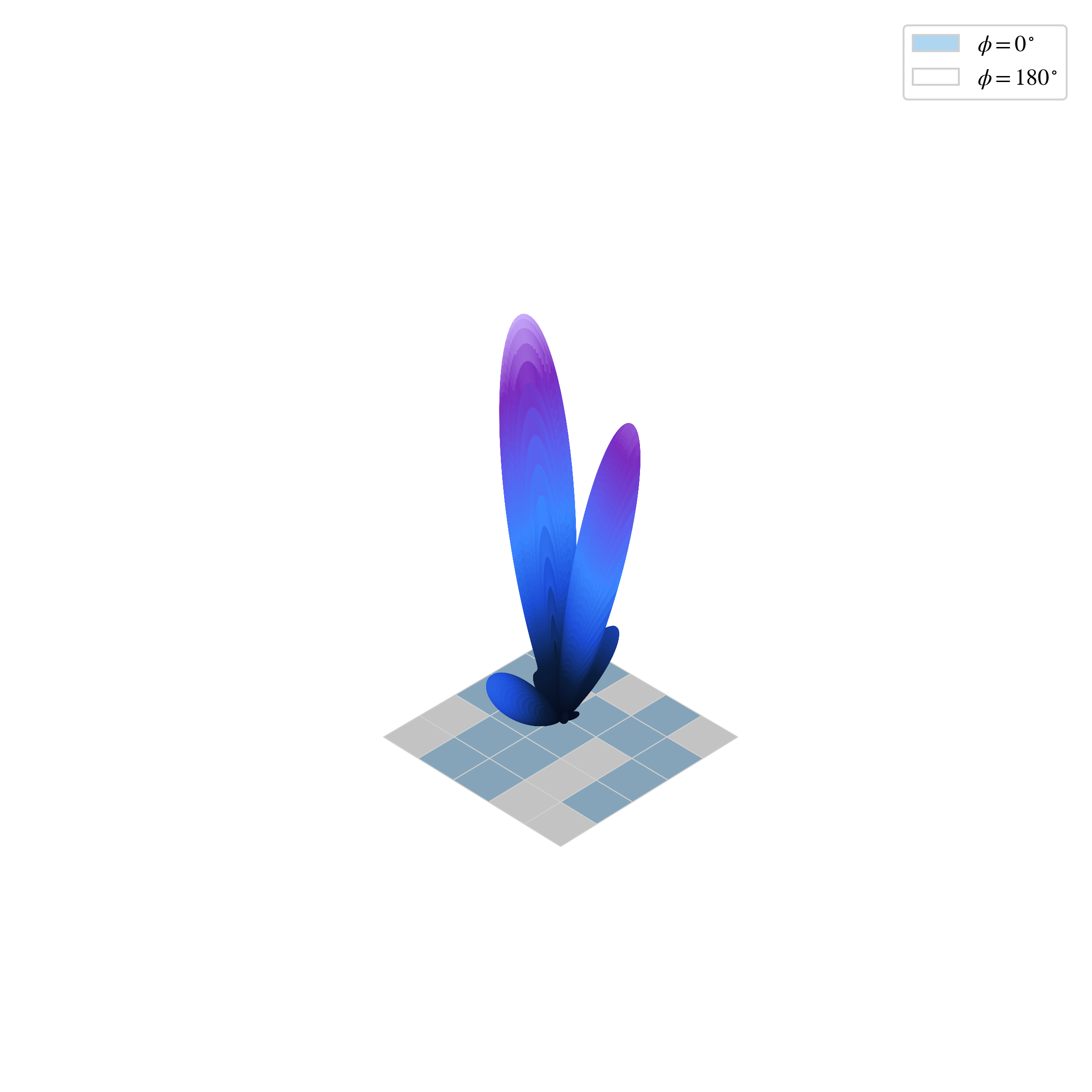}%
    \hspace{-4mm}%
    \includegraphics[width=0.131\textwidth, keepaspectratio=true, clip=true, trim=6.6cm 4.5cm 5.4cm 5.5cm]{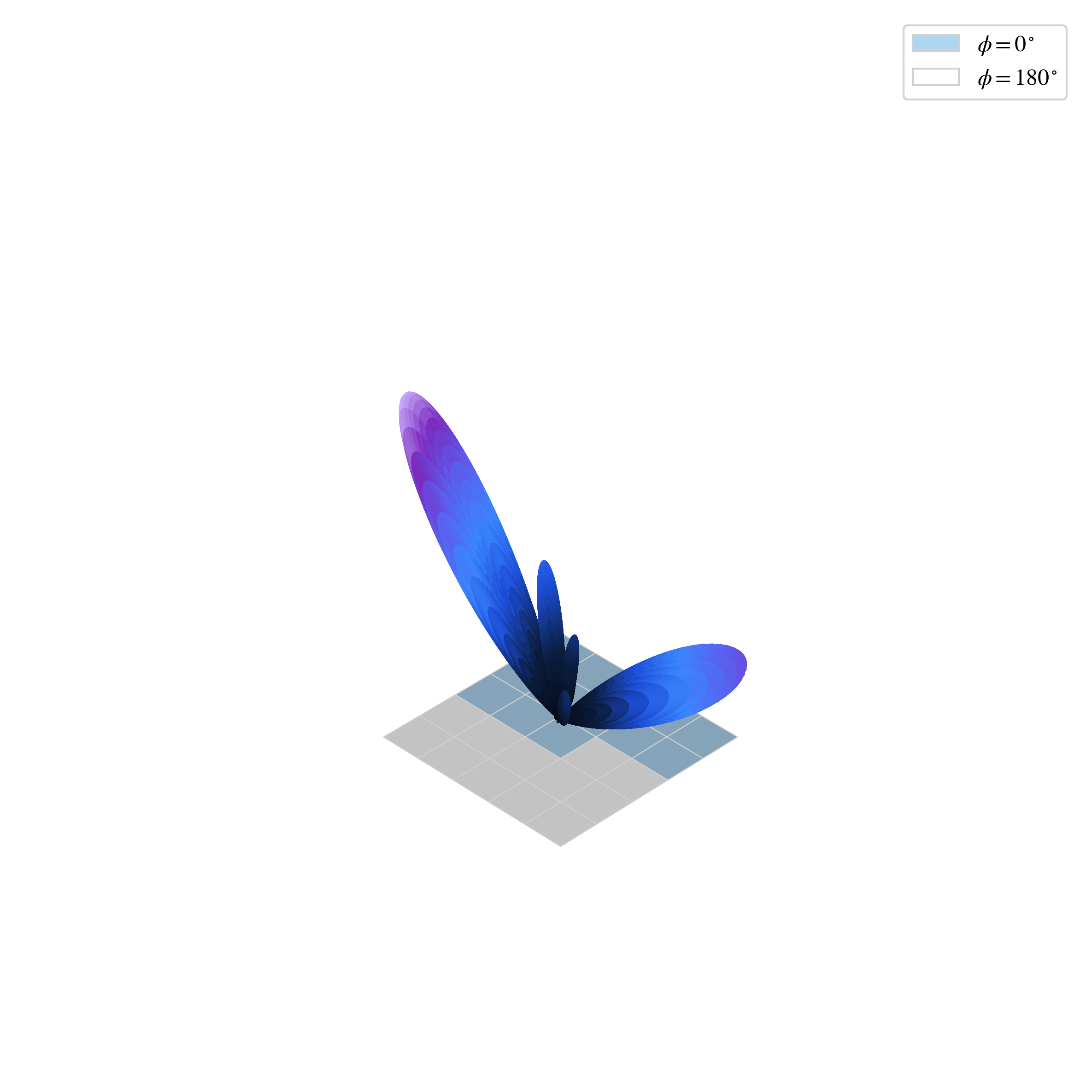}%
    \hspace{-4mm}%
    \includegraphics[width=0.131\textwidth, keepaspectratio=true, clip=true, trim=6.6cm 4.5cm 5.4cm 5.5cm]{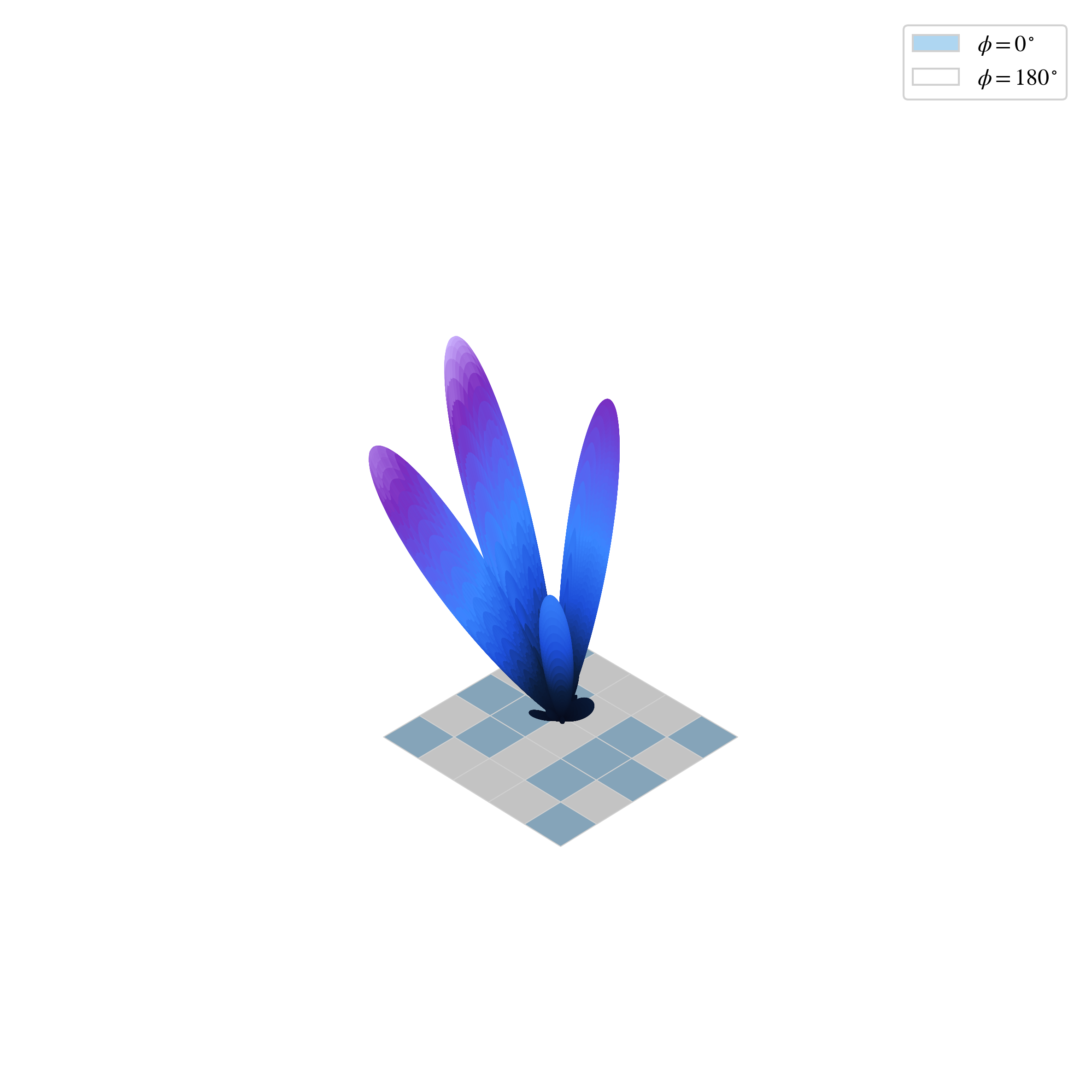}%
    }
    \caption{Overview of QAOA-based reconfigurable intelligent surface (RIS) optimization. The QAOA circuit optimizes binary phase configurations of the RIS, yielding different radiation patterns.}
    \label{fig:teaser}
\end{teaserfigure}

\maketitle

\section{Introduction}
Reconfigurable Intelligent Surfaces (RIS) have emerged as a transformative technology in modern electromagnetics and are expected to play a foundational role in future 6G communication systems~\cite{rana_review_2023}. These metasurfaces consist of numerous controllable elements that shape and manipulate reflected electromagnetic waves, enabling wavefront engineering and adaptive signal redirection. RIS design consists of determining the optimal configuration of discrete phase states across the array to achieve a desired radiation pattern. As array sizes grow, this task becomes a high-dimensional combinatorial optimization problem. Classical approaches typically rely on heuristics or relaxation methods, which can struggle with scalability due to the exponential growth of the solution space. This has motivated the exploration of using quantum optimization algorithms for RIS optimization~\cite{batool_quantum-enhanced_2026}. Among these, gate-based methods, and in particular the Quantum Approximate Optimization Algorithm (QAOA)~\cite{farhi_quantum_2014,choi_tutorial_2019}, see Fig.~\ref{fig:teaser}, offer a flexible framework for encoding discrete optimization problems while retaining compatibility with near-term quantum hardware~\cite{colella_quantum_2024,ross_engineering_2022}.

A key limitation in existing quantum formulations of RIS optimization is the reliance on simplified models that neglect important electromagnetic effects. In practical systems, physical constraints such as sub-wavelength element spacing ($d \le \lambda/2$) introduce strong mutual coupling between elements. This coupling alters the effective response of the array and invalidates the common assumption of independent radiators, leading to discrepancies between optimized configurations and actual engineering performance.

In this work, we investigate how increasing physical realism in RIS modeling impacts its suitability for quantum optimization. Specifically, we construct and analyze a set of Ising interaction matrices ($J_{ij}$) that incorporate varying levels of electromagnetic effects, from simplified phase-target models to more physically grounded formulations. By examining these representations within the QAOA framework, we aim to clarify the trade-offs between model fidelity and quantum hardware feasibility, and to identify practical pathways for implementing RIS optimization on near-term quantum devices~\cite{lee_quantum_2025}. Notably, these models are not intended as deployment-level descriptions of RIS, but rather as physics-informed toy models that retain sufficient simplicity to be mapped to QUBO instances while extending beyond the standard independent-element formulation. We also leave detailed noise effects outside the scope of this work, focusing instead on the impact of increasing model realism.

This paper is organized as follows: Section~\ref{sec:background-ris} derives the RIS electromagnetic objective and coupling structure, Section~\ref{sec:background-QUBO-opt-w-QAOA} introduces QAOA and then maps the RIS objective into the necessary QUBO formalism, Section~\ref{sec:methodology} details the methodology, and Section~\ref{sec:results} reports numerical results.

\section{Background} 

\subsection{Reconfigurable Intelligent Surfaces (RIS)}

\label{sec:background-ris}

Reconfigurable Intelligent Surfaces (RIS) are metasurfaces composed of many elements used to shape the reflection of an electromagnetic wave, enabling not only signal redirection but also wavefront manipulation. They are currently under active development and expected to play a foundational role in future 6G systems. 

The individual elements typically have sizes comparable to the wavelength, usually ranging from $\lambda/10$ to $\lambda/2$. The advantages of RIS are commonly attributed to their ease of deployment, environmental friendliness, compatibility with different electromagnetic and communication systems, and their ability to enhance spectral and power efficiency, especially over long distances~\cite{liu_reconfigurable_2021,elmossallamy_reconfigurable_2020}.

These metasurfaces are reconfigured by adjusting the local reflection phase of each element. For simplicity, and to map the physical system to a discrete optimization problem, we consider a binary phase configuration where each element can apply a phase shift of $0$ or $\pi$. This naturally corresponds to reflection coefficients of $+1$ and $-1$, respectively. 

\begin{figure}[t]
    \centering
    \includegraphics[width=0.99\linewidth]{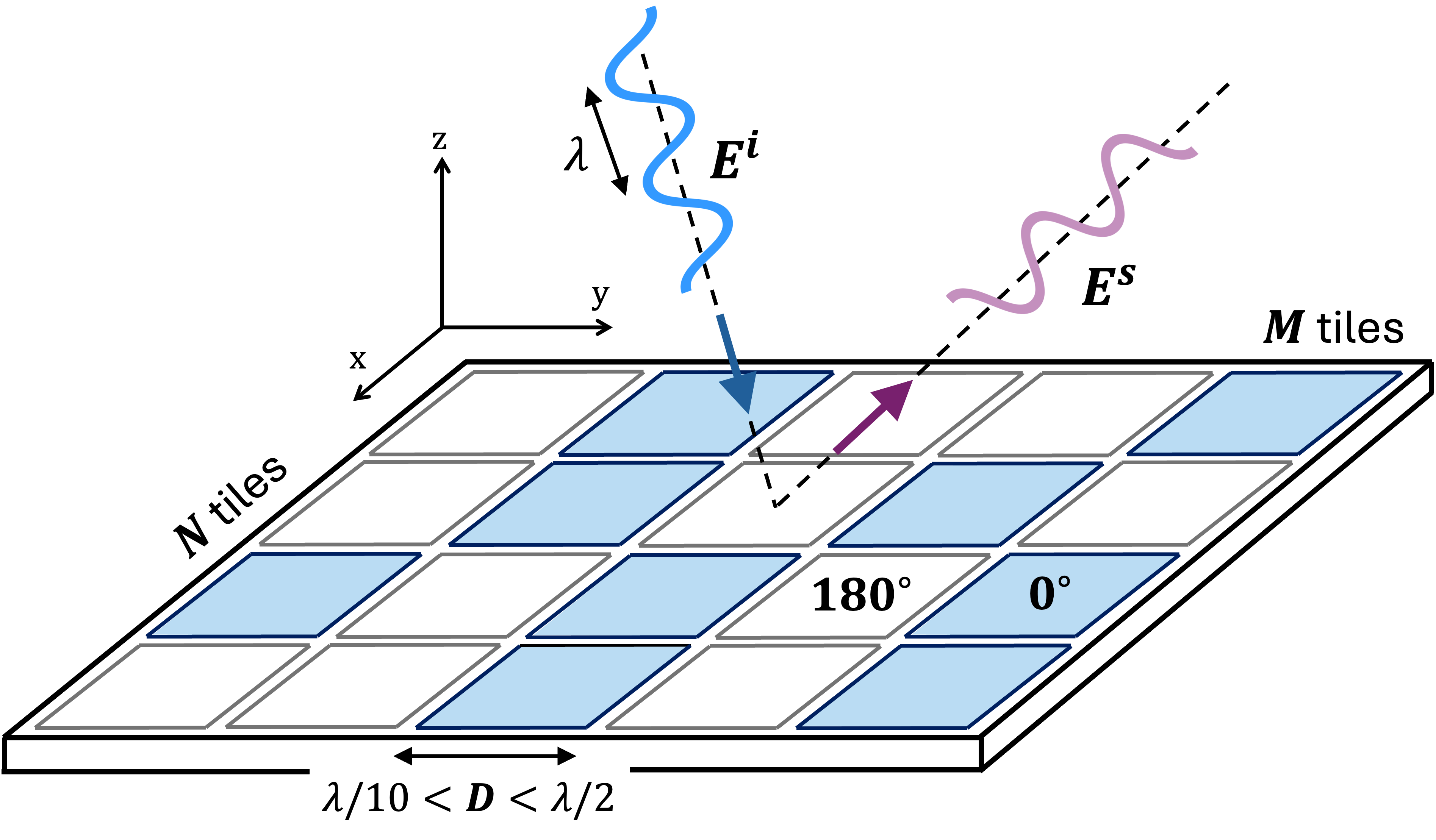}

    \vspace{-2mm}
    \caption{Illustration of a reconfigurable intelligent surface (RIS) reflecting an incident wave toward a desired target direction, given some phase shift for each tile. Here, the incident and scattered fields are assumed to be in the far field.}
    \label{fig:ris_concept}
\end{figure}

We consider a metasurface composed of $M \times N$ elements, shown in Fig.~\ref{fig:ris_concept}, which for simplicity can be assumed to be a square grid. In practice, RIS structures may consist of arrays on the order of $10 \times 10$ elements, depending on the application and operating frequency.

The primary objective in RIS design is to determine the optimal configuration of element phases such that an incoming wave from an incident direction $(\theta^i, \phi^i)$ is reflected toward a target direction $(\theta^s, \phi^s)$. Using the physical optics approximation, this beam-steering problem can be formulated as maximizing the radar cross section (RCS)~\cite{lee_quantum_2026}, denoted by $\sigma$, integrated over a specific target solid angle $\Omega_{target}$:
\begin{equation}
\max_{\{x\}} \iint_{\Omega_{target}} \sigma(\theta^s, \phi^s) \, d\Omega^s.
\label{eq:optimization-objective}
\end{equation}
The RCS in any arbitrary observation direction $(\theta^s, \phi^s)$ is proportional to the scattered power intensity. This intensity can be factored into the individual element scattering pattern $F_e$ and the macroscopic array factor $AF$:
\begin{equation*}
\sigma(\theta^s, \phi^s) \propto \left| F_e(\theta^s, \phi^s, \theta^i) \right|^2 \; \left| AF(\theta^s, \phi^s, \theta^i) \right|^2.
\end{equation*}
Assuming a uniform element spacing of $d$, and denoting the binary phase state of each element as $x \in \{-1,+1\}$, the squared magnitude of the array factor can be expanded as a double summation over all element pairs $(p,q)$ and $(u,v)$ in the $M \times N$ grid:
\begin{equation*}
\left| AF \right|^2 = \sum_{p=1}^{M} \sum_{q=1}^{N} \sum_{u=1}^{M} \sum_{v=1}^{N} x_{pq} x_{uv} \exp\left( j \left[ k_x (p-u)d + k_y (q-v)d \right] \right),
\end{equation*}
where $k_x$ and $k_y$ represent the spatial frequencies dictated by the incoming and scattered wave vectors.

Substituting this expansion back into the optimization integral, and noting that the imaginary sine components cancel out due to the geometric symmetry of the array, we can isolate a real-valued physical interaction weight $W_{pq,uv}$ for any pair of elements:
\begin{equation}
W_{pq,uv} = \iint_{\Omega_{target}} \left| F_e \right|^2 \cos\!\left( k_x (p-u)d + k_y (q-v)d \right) d\Omega^s.
\label{eq:W-matrix-entries}
\end{equation}
While this formulation provides a clean mathematical bridge to quantum devices (see Section~\ref{sec:background-QUBO-opt-w-QAOA}), it relies heavily on the idealized assumption that the total array factor is a simple geometric superposition of isolated radiators. In physical deployments, preventing the formation of grating lobes requires an element spacing of $d \le \lambda/2$. At these tight proximities, strong electromagnetic mutual coupling occurs between the metallic elements, heavily distorting the local phase responses and invalidating the standard array factor approximation. Consequently, optimizing $\mathbf{W}$ alone is insufficient for accurate beam steering in real hardware, requiring models that account for physical coupling.

\subsection{RIS optimization using QAOA}

\label{sec:background-QUBO-opt-w-QAOA}

The \textit{quantum approximate optimization algorithm} (QAOA) is a hybrid quantum-classical algorithm designed to solve combinatorial optimization problems on quantum hardware~\cite{farhi_quantum_2014}. In particular, QAOA targets problems formulated as \textit{quadratic unconstrained binary optimization} (QUBO) problems~\cite{punnen_quadratic_2022}. Formally, a QUBO instance is defined as finding the vector $\mathbf{z}^*=(z_1,...,z_N)^T$ of binary entries\footnote{In the literature, QUBOs are often defined using binary variables $x_i \in\{0, 1\}$.} $z_i\in\{\pm1\}$ satisfying
\begin{equation}
    \mathbf{z}^* = \arg\min_{\mathbf{z}} \mathbf{z}^T\mathbf{J}\mathbf{z} + \mathbf{h}^T\mathbf{z}.
    \label{eq:Ising_formulation}
\end{equation}
Here, $\mathbf{J}\in \mathbb{R}^{N\times N}$ is a symmetric matrix describing the strength of pairwise couplings between binary variables, while $\mathbf{h}\in\mathbb{R}^N$ contains linear bias terms. 
A QUBO instance can be mapped to a quantum ground-state search problem by associating each binary variable $z_i$ with the Pauli-$Z$ operator $Z_i$ acting on qubit $i$. Hence, we assume that the number of qubits matches the number of binary variables. The problem then translates to finding the quantum state vector $\ket{\psi_\text{gs}}$ that satisfies
\begin{equation*}
    \ket{\psi_\text{gs}} = \arg\min_{\ket{\psi}} \braket{\psi | H_C |\psi},
\end{equation*}
where the \textit{cost Hamiltonian} $H_C$ is defined as
\begin{equation*}
    H_C = \sum_{\langle i,j \rangle} J_{i,j} Z_i Z_j + \sum_i h_i Z_i,
\end{equation*}
where $\braket{i,j}$ denotes the set of pairs $(i,j)$ for which $J_{i,j} \neq 0$.

The working principle of QAOA is to approximate $\ket{\psi_\text{gs}}$ using a parametrized quantum circuit
\begin{equation*}
    \ket{\psi(\boldsymbol{\gamma},\boldsymbol{\beta})} =
    \prod_{l=1}^{p} 
    \underbrace{e^{-i \beta_l H_M} e^{-i \gamma_l H_C}}_{U_l(\beta_l,\gamma_l)}
    \ket{\psi_0}.
\end{equation*}
where $p$ denotes the circuit depth and the parameters $\gamma_l$ and $\beta_l$ are variational angles of the $l$-th layer of the QAOA ansatz~\cite{farhi_quantum_2014}. The initial state is set to the product state $\ket{\psi_0} = \ket{+}^{\otimes N}$. In the standard QAOA formulation, the \textit{mixer Hamiltonian} $H_M$ is typically chosen as
\begin{equation*}
    H_M = \sum_i X_i,
\end{equation*}
with $X_i$ denoting the Pauli-$X$ operator on qubit $i$.

QAOA is a \textit{hybrid quantum-classical} algorithm, as the trainable parameters $\boldsymbol{\gamma},\boldsymbol{\beta}$ are optimized using a classical feedback loop, with the objective being to minimize the energy expectation value $\langle \psi(\boldsymbol{\gamma},\boldsymbol{\beta})|H_C| \psi(\boldsymbol{\gamma},\boldsymbol{\beta})\rangle$. Upon convergence of the parameter optimization, samples from the optimized quantum state can be taken as candidate solutions to the optimization problem.

Crucially, the RIS objective in Eq.~\eqref{eq:optimization-objective} can be directly mapped into a QUBO formulation compatible with QAOA optimization. By vectorizing the $M \times N$ binary phase grid into a vector $\mathbf{x} \in \{-1,+1\}^{MN}$, the metasurface optimization objective can be reformulated as
\begin{equation*}
\max_{\mathbf{x}\in\{-1,+1\}^{MN}} \mathbf{x}^{T}\mathbf{W}\mathbf{x}
\; \Longleftrightarrow \; \min_{\mathbf{x}\in\{-1,+1\}^{MN}}\mathbf{x}^{T}(-\mathbf{W})\mathbf{x}
\end{equation*}
where the interaction matrix entries are defined by Eq.~\eqref{eq:W-matrix-entries}. This formulation matches the QUBO definition of Eq.~\eqref{eq:Ising_formulation} with coupling matrix $\mathbf{J}=-\mathbf{W}$ and linear coefficients $\mathbf{h}=\mathbf{0}$.

\section{Methodology}
\label{sec:methodology}
\begin{figure*}[h]
    \centering
    \includegraphics[width=0.95\textwidth, clip = true, trim = 0cm 0 0 0.4cm]{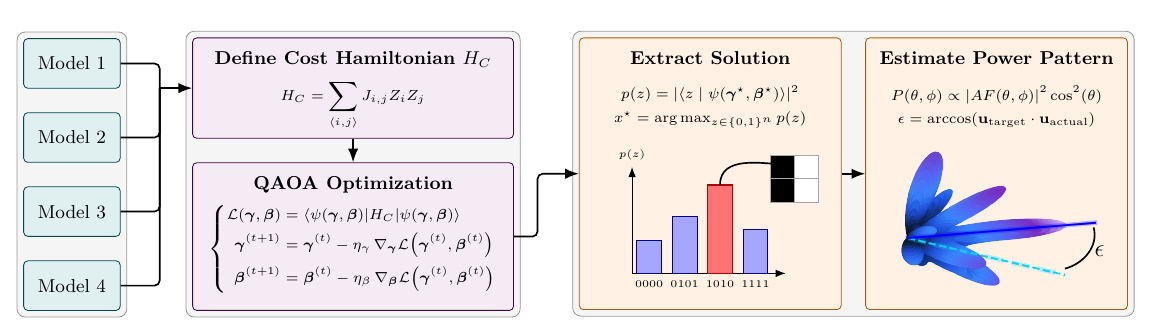}

    \vspace{-2mm}
    \caption{Workflow for QAOA-based metasurface optimization. The four models feed into the cost-Hamiltonian construction which is optimized using QAOA. The optimized circuit parameters are then used to extract the most likely binary solution and estimate the radiation power pattern. The final performance is summarized by the angular mismatch metric $\epsilon$. }
    \label{fig:workflow}
\end{figure*}

To solve the energy minimization problem, we first define the system parameters: the operating frequency $f$ (from which the free-space wavelength $\lambda$ is derived), the spatial element size, the grid dimensions ($N \times N$), the depth of the QAOA circuit ($p$), and the incident and scattered wave directions, denoted as $(\theta^i,\phi^{i})$ and $(\theta^{s},\phi^{s})$, respectively. Next, we construct the Ising interaction matrix, $J_{ij}$, which governs the pairwise coupling between qubits. This matrix is designed to capture the underlying electromagnetic interactions and the spatial relationships between the RIS elements. The explicit formulation of $J_{ij}$ is detailed in the following section.

Once $J_{ij}$ is defined, we construct the cost Hamiltonian $H_C$ and the transverse-field mixer Hamiltonian $H_M$. These are applied alternately to form a single layer of the QAOA circuit, which is repeated for a total depth of $p$. The quantum circuit is simulated using a custom state-vector simulator~\cite{asadi_hybrid_2024} implemented in PyTorch. 

For the classical optimization loop, various optimizers (e.g., L-BFGS-B, Adam, COBYLA, SPSA) can be employed, each with varying stability and convergence properties. For parameter optimization, we utilize the gradient-based Adam optimizer~\cite{kingma_adam_2017}, with exact gradients computed via adjoint differentiation. At each iteration, the optimizer updates the variational parameters $\boldsymbol{\gamma}$ and $\boldsymbol{\beta}$ (representing the evolution times for $H_C$ and $H_M$, respectively) to navigate the energy landscape toward a local minimum. 

To avoid converging to suboptimal local minima, the optimization can be restarted multiple times from different initial conditions. Specifically, we initialize the $\boldsymbol{\gamma}$ and $\boldsymbol{\beta}$ parameters using a linear progression across the layers, perturbed by randomized Gaussian noise. This multi-start approach can be crucial, as sampling different regions of the complex optimization landscape can increase the probability of finding the global minimum.

To extract the final solution from the optimized circuit, two approaches exist: finite measurement sampling (which mimics actual quantum hardware and introduces statistical noise) or reading the exact statevector probabilities. We utilize the latter to determine the optimal binary string, bypassing sampling noise entirely. The resulting binary quantum states are then mapped back to the physical configurations of the RIS tiles (e.g., mapping state $|0\rangle \to 0$ rad, and $|1\rangle \to \pi$ rad). For smaller RIS sizes (e.g., $5 \times 5$), we additionally perform an exhaustive search over all $2^{N^2}$ configurations to identify the true global optimum (including degenerate solutions). This provides a deterministic baseline for evaluating QAOA performance and assessing how well the QUBO models reflect the underlying physics. Benchmarking against scalable classical heuristics remains important for assessing performance at larger array sizes.

Finally, to validate the accuracy of the optimized RIS configuration, we simulate the scattered far-field radiation pattern. We compute the coupled array factor using a spherical wave mutual coupling model to evaluate the true physical beamforming performance. The far-field array factor $AF(\theta, \phi)$~\cite{balanis_balanis_2023} is calculated as:
\begin{equation*}
    AF(\theta, \phi) = \sum_{i=1}^{N_{total}} v_{\text{actual},i} \exp\left[ j k \left( x_i \sin\theta \cos\phi + y_i \sin\theta \sin\phi \right) \right],
\end{equation*}
where $N_{total}$ is the total number of elements, $k$ is the wavenumber, and $(x_i, y_i)$ are the spatial coordinates of the $i$-th element. The complex excitation of the $i$-th element is given by $v_{\text{actual},i}$, which is an element of the coupled excitation vector $\mathbf{v}_{\text{actual}} = \mathbf{C} \, \mathbf{v}_{\text{ideal}}$. The ideal phase excitations are defined as $v_{\text{ideal},i} = e^{j(\Phi_{\text{RIS},i} - \Phi_{\text{in},i})}$, where $\Phi_{\text{RIS},i}$ is the optimized phase state and $\Phi_{\text{in},i}$ is the incident phase. The mutual coupling matrix $\mathbf{C}$ models spherical wave decay, defined as $C_{ij} = \alpha \frac{\cos(k d_{ij})}{d_{ij}}$ for $i \neq j$ (with $\alpha$ being the coupling strength coefficient and $d_{ij}$ the physical distance between elements $i$ and $j$), and self-coupling is assumed ideal ($C_{ii} = 1$).

To quantify the performance of the Ising model and the optimization steps, we compute the spatial pointing error. The normalized power pattern, computed as the product of the array factor squared and the element pattern (e.g., $P(\theta, \phi) \propto |AF(\theta, \phi)|^2 \cos^2\theta$), is evaluated over a dense angular grid to identify the actual peak radiation direction $(\theta_{\text{actual}}, \phi_{\text{actual}})$. Both the target and actual pointing directions are converted into 3D Cartesian unit vectors, $\mathbf{u}_{\text{target}}$ and $\mathbf{u}_{\text{actual}}$, using standard spherical-to-Cartesian transformations (e.g., $\mathbf{u} = [\sin\theta \cos\phi, \sin\theta \sin\phi, \cos\theta]^T$). 

The total 3D spatial error, $\epsilon$, is then defined as the angle between these two unit vectors, calculated via the inverse cosine of their dot product:
\begin{equation*}
    \epsilon = \arccos\left(\mathbf{u}_{\text{target}} \cdot \mathbf{u}_{\text{actual}}\right).
\end{equation*}
The complete optimization and validation pipeline, from the initial cost-Hamiltonian construction to the final extraction of this angular mismatch metric $\epsilon$, is visually summarized in Fig.~\ref{fig:workflow}.

\subsection{Formulating the Coupling Matrix $J_{ij}$}
\begin{table*}[h]
\centering
\caption{Comparison of $J_{ij}$ construction models.}

\label{tab:qubo_models}

\renewcommand{\arraystretch}{1.3}
\setlength{\tabcolsep}{12pt}

\vspace{-3mm}
\begin{tabular}{clll}
\toprule
\textbf{Model} & \textbf{QUBO Entries} & \textbf{Physical Fidelity} & \textbf{Quantum Suitability} \\
\midrule

\textbf{1} 
& {\small $J_{ij} = \cos(\Delta \phi_{ij})$} 
& Poor (no coupling) 
& Poor (fully dense) \\

\textbf{2} 
& {\small $\displaystyle 
J_{ij} =
\begin{cases}
\cos(\Delta \phi_{ij}) + \dfrac{k}{d_{ij}}, & d_{ij} < d_\text{cutoff} \\
0, \; \; \;  \text{otherwise} & 
\end{cases}$} 
& Approximate (no wave physics) 
& Good (sparse) \\

\textbf{3} 
& {\small $\displaystyle 
J_{ij} =
\begin{cases}
\cos(\Delta \phi_{ij}) + \alpha \dfrac{\cos(k d_{ij})}{d_{ij}}, & d_{ij} < d_\text{cutoff} \\
0, \; \; \;  \text{otherwise} & 
\end{cases}$} 
& Good (includes physical coupling) 
& Good (sparse) \\

\textbf{4} 
& {\small $J_{ij} = \mathrm{Re}(B_i B_j^*)$} 
& Optimal (full physics) 
& Poor (fully dense) \\

\bottomrule
\end{tabular}
\end{table*}

To construct the Ising interaction matrix $J_{ij}$, four distinct models were evaluated, each presenting a different trade-off between electromagnetic accuracy and quantum hardware feasibility (Table~\ref{tab:qubo_models}). 

First, let us define the ideal phase relations based on the physical geometry of the array. Let $(x_i, y_i)$ denote the physical coordinates of the $i$-th element on the RIS grid, and $k = 2\pi/\lambda$ represent the free-space wavenumber. For an incident wave arriving from angles $(\theta^i, \phi^i)$ and a desired target scattered wave pointing toward $(\theta^s, \phi^s)$, the spatial phases at element $i$ are given by the projection of the wave vectors onto the array plane:
\begin{equation*}
\begin{cases}
    \phi^{in}_i  &= k \left( x_i \sin\theta^i \cos\phi^i + y_i \sin\theta^i \sin\phi^i \right) \\
\phi^{out}_i &= k \left( x_i \sin\theta^s \cos\phi^s + y_i \sin\theta^s \sin\phi^s \right) .
\end{cases}
\end{equation*}
The ideal phase shift required at element $i$ to steer the beam is the difference between the outgoing and incoming spatial phases, $\Phi_i = \phi^{out}_i - \phi^{in}_i$. Consequently, the ideal target phase difference between any two elements $i$ and $j$ is defined as:
\begin{equation*}
\Delta\phi_{ij} = \Phi_i - \Phi_j = (\phi^{out}_i - \phi^{in}_i) - (\phi^{out}_j - \phi^{in}_j).
\end{equation*}
Additionally, let $d_{ij} = \sqrt{(x_i - x_j)^2 + (y_i - y_j)^2}$ represent the physical Euclidean distance separating elements $i$ and $j$. Using these spatial phase differences as a baseline, we construct the following four coupling models.

\subsubsection{Model 1: Ideal Dense Phase Coupling}
This baseline model defines the interactions based purely on the idealized phase difference target, without considering physical separation:
\begin{equation*}
J_{ij} = \cos(\Delta \phi_{ij}).
\end{equation*}
From a physical perspective, this model is fundamentally flawed because it ignores physical mutual coupling entirely. While the elements are mathematically forced to interfere constructively in the far-field, the model assumes perfect hardware isolation, which is unrealistic for closely spaced RIS elements. From a quantum perspective, it is equally problematic. The resulting Hamiltonian is fully dense, requiring all-to-all logical qubit interactions to implement the time evolution generated by $H_C$. Mapping such interactions onto current superconducting quantum hardware, which is constrained by planar layouts with limited qubit connectivity, incurs substantial overhead due to qubit routing. This, in turn, increases gate counts and circuit depth, leading to worsened decoherence and accumulated errors.

\subsubsection{Model 2: Sparse Distance-Penalized Coupling}
This model adds a generic inverse-distance penalty (scaled by a constant $C$) to the phase differences, applying a spatial cutoff to sever interactions beyond adjacent elements:
\begin{equation*}
J_{ij} = \begin{cases} \cos(\Delta \phi_{ij}) + \frac{k}{d_{ij}} & \text{if } d_{ij} < d_{\text{cutoff}} \\ 0 & \text{otherwise} \end{cases}.
\end{equation*}
Physically, this constitutes an approximation. It acknowledges that coupling decays with distance, but the generic $1/d$ penalty lacks rigorous wave physics, primarily because it omits wavelength-dependent phase variation. However, from a quantum computing perspective, this model is preferable as the spatial cutoff ensures that the $J_{ij}$ matrix remains sparse. This maps exceptionally well to the planar topologies of real superconducting quantum processors, requiring far fewer qubit routing operations and enabling the execution of QAOA at significantly greater circuit depths.

\subsubsection{Model 3: Sparse Spherical Wave Coupling}
Here we propose an improvement upon Model 2 by using the real part of a spherical wave function (scaled by a coefficient $\alpha$) for mutual coupling, while maintaining the strict spatial cutoff:
\begin{equation*}
J_{ij} = \begin{cases} \cos(\Delta \phi_{ij}) + \alpha \frac{\cos(k d_{ij})}{d_{ij}} & \text{if } d_{ij} < d_{\text{cutoff}} \\ 0 & \text{otherwise} \end{cases}.
\end{equation*}
From a physical standpoint, this model is a significant step forward. It introduces a physically meaningful oscillating behavior, allowing for both constructive and destructive local coupling depending on element separation. From a quantum perspective, it remains highly efficient due to the sparsity of the Hamiltonian.

\subsubsection{Model 4: Dense Corrected QUBO Mapping}
To rigorously map a coupled RIS to an Ising Hamiltonian, we derive the interactions directly from the far-field objective. The scattered electric field in the target direction is expressed as the projection of the actual element excitations onto the steering vector $\mathbf{V}_{out}$, where each excitation accounts for mutual coupling, the incident field, and the binary phase state.
\begin{equation*}
E = \sum_{i=1}^{N^2} V_{out,i} \, v_{\text{actual},i}
\quad \text{where} \quad
v_{\text{actual},i} = \sum_{j=1}^{N^2} C_{ij} \, x_j \, V_{in,j}^*.
\end{equation*}
Note that $V_{in}^* = \exp(-j \phi^{in})$ represents the phase shift required to compensate for the incoming wave's propagation delay. Crucially, unlike Model 3, the mutual coupling matrix here employs the full complex scalar free-space Green's function, $C_{ij} = \alpha \frac{\exp(-j k d_{ij})}{d_{ij}}$ (with $C_{ii} = 1$), ensuring both active and reactive near-field power flows are captured. Substituting this physical excitation back into the total electric field equation yields:
\begin{equation*}
E = \sum_{i=1}^{N^2} V_{out, i} \left( \sum_{j=1}^{N^2} C_{ij} x_j V_{in, j}^* \right).
\end{equation*}
Because the mutual coupling matrix for a passive array is symmetric ($C_{ij} = C_{ji}$) due to Lorentz reciprocity, we can rearrange the summations to isolate the binary optimization variable $x_j$:
\begin{equation*}
E = \sum_{j=1}^{N^2} x_j \left( V_{in, j}^* \sum_{i=1}^{N^2} C_{ij} V_{out, i} \right) = \sum_{j=1}^{N^2} B_j x_j.
\end{equation*}
We encapsulate these physical constraints regarding wave incidence, the complex mutual coupling, and the target pattern, into a single effective complex vector $\mathbf{B} \in \mathbb{C}^N$, with entries:
\begin{equation*}
B_j = V_{in, j}^* \sum_{i=1}^{N^2} C_{ij} V_{out, i}.
\end{equation*}
Given that quantum optimizers operate by minimizing or maximizing an energy landscape, we map this directly to power. In electromagnetics, radiated power $P$ is proportional to the square of the electric field magnitude ($|E|^2 = E \cdot E^*$). Since our quantum variables are binary phase states $x \in \{-1, 1\}$, they are strictly real, meaning $x_j = x_j^*$. The expansion thus perfectly matches a standard Quadratic Unconstrained Binary Optimization (QUBO) form:
\begin{equation*}
|E|^2 = \left( \sum_{i=1}^{N^2} B_i x_i \right) \left( \sum_{j=1}^{N^2} B_j^* x_j^* \right) = \sum_{i=1}^{N^2} \sum_{j=1}^{N^2} (B_i B_j^*) x_i x_j.
\end{equation*}
Finally, we extract the Ising interaction matrix ($J_{ij}$). The Ising Hamiltonian requires real-valued interaction weights. For any pair of antennas $i$ and $j$, the sum includes both the $(i,j)$ and $(j,i)$ terms:
\begin{equation*}
B_i B_j^* + B_j B_i^* = 2 \text{Re}(B_i B_j^*).
\end{equation*}
Because the imaginary components cancel out, the interaction weight $J_{ij}$ required for the off-diagonal terms of the quantum Hamiltonian is simply the real part of the outer product of $B$:
\begin{equation*}
J_{ij} = \text{Re}(B_i B_j^*).
\end{equation*}
By maximizing this specific Hamiltonian $\sum J_{ij} x_i x_j$, the quantum algorithm inherently finds the bitstring $x$ that mathematically pre-compensates for the complex physical coupling matrix $C$, thereby maximizing the actual power in the target direction. 

From a physical perspective, this is the optimal approach, as it perfectly compensates for global array coupling without relying on arbitrary distance cutoffs. However, from a quantum execution perspective, removing the sparsity constraint results in a fully dense matrix. While mathematically elegant and easily solved on state-vector simulators, running this on physical quantum hardware poses severe difficulties due to the massive overhead required for qubit routing operations.

Since the four coupling models produce $J_{i,j}$ values on very different scales, we run QAOA on a globally rescaled coupling matrix, $\tilde{J}_{i,j}=J_{i,j}/\sum_{i<j}\lvert J_{i,j }\rvert$, where each interaction is normalized by the total absolute weight of all unique pairwise couplings. This fixes the overall interaction-strength scale across models, hence improving convergence behavior while preserving the optimal solution set.

\section{Physics-Based RIS Modeling Results}
\label{sec:results}

\begin{figure}[b]
    \noindent 
    \begin{minipage}[t]{\columnwidth}
        \noindent\rule{\columnwidth}{0.4pt}
        
        \vspace{1mm}
        \textbf{Note:} \textit{One must distinguish between the simplified model used for quantum optimization and the full physical model used for validation. We first construct an approximate coupling model that maps RIS electromagnetic interactions to a QUBO, defining the Ising matrix ($J_{ij}$) explored by QAOA. \textbf{However, reaching the QUBO optimum does not guarantee optimal electromagnetic performance}. The resulting phase configuration is therefore evaluated using a fully coupled electromagnetic model to compute the true far-field response, allowing a direct comparison and assessment of how accurately the quantum model reflects real beamforming performance.}
        
        \vspace{-0.7mm}
        \noindent\rule{\columnwidth}{0.4pt}
    \end{minipage}
\end{figure}

\begin{figure*}[t]
    \centering

    \raisebox{-0.0\height}{
    \begin{minipage}[b]{0.485\textwidth}
    \centering
    
    \begin{minipage}[b]{0.435\linewidth}
        \centering
        \includegraphics[width=\linewidth, clip=true, trim = 0 1.7cm 0 0]{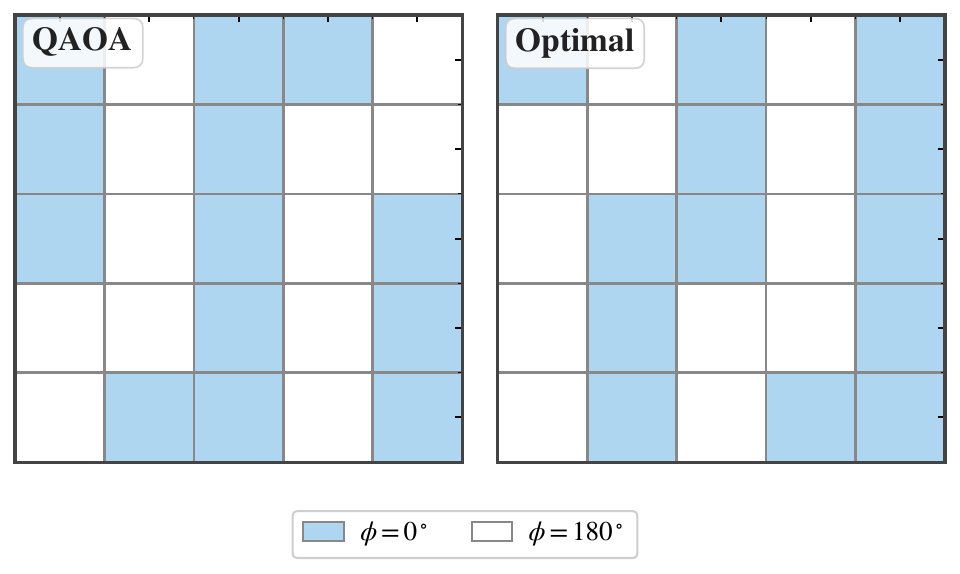}\\[0pt]
        \makebox[0.5\linewidth][c]{$\epsilon = 2.23^\circ$}%
        \makebox[0.5\linewidth][c]{$\epsilon = 3.66^\circ$}\\[4pt]
        (a) Model 1
    \end{minipage}
    \hspace{7mm}
    \begin{minipage}[b]{0.435\linewidth}
        \centering
        \includegraphics[width=\linewidth, clip=true, trim = 0 1.7cm 0 0]{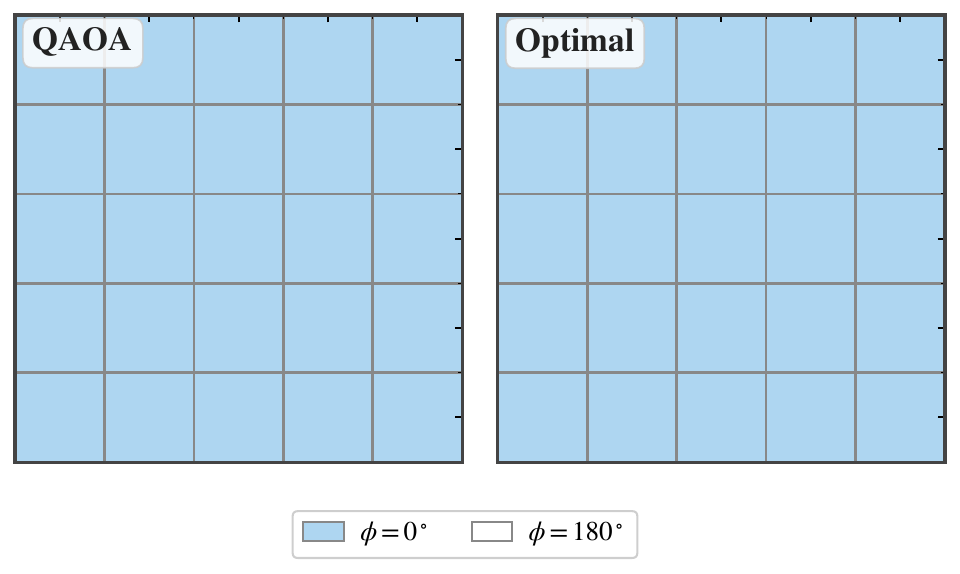}\\[0pt]
        \makebox[0.5\linewidth][c]{$\epsilon = 33.14^\circ$}%
        \makebox[0.5\linewidth][c]{$\epsilon = 33.14^\circ$}\\[4pt]
        (b) Model 2
    \end{minipage}

    \vspace{-1mm}
    
    \begin{minipage}[b]{\linewidth}
        \centering
        \includegraphics[width=0.35\linewidth, clip=true, trim = 4.7cm 0 5.4cm 8.6cm]{images/RIS_2D_Comparison2.pdf}
    \end{minipage}
    
    \vspace{-0mm}
    
    \begin{minipage}[b]{0.435\linewidth}
        \centering
        \includegraphics[width=\linewidth, clip=true, trim = 0 1.7cm 0 0]{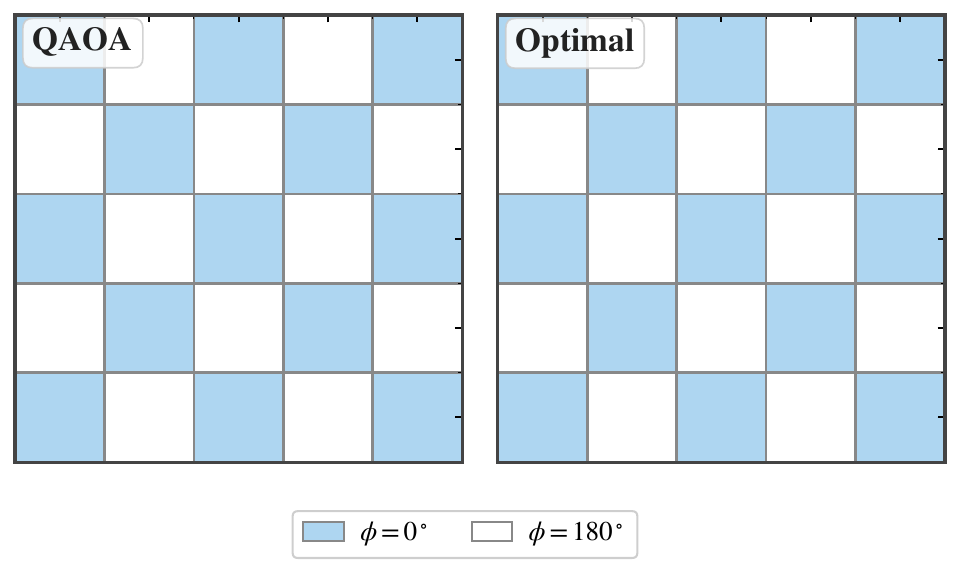}\\[0pt]
        \makebox[0.5\linewidth][c]{$\epsilon = 45.96^\circ$}%
        \makebox[0.5\linewidth][c]{$\epsilon = 45.96^\circ$}\\[4pt]
        (c) Model 3
    \end{minipage}
    \hspace{7mm}
    \begin{minipage}[b]{0.435\linewidth}
        \centering
        \includegraphics[width=\linewidth, clip=true, trim = 0 1.7cm 0 0]{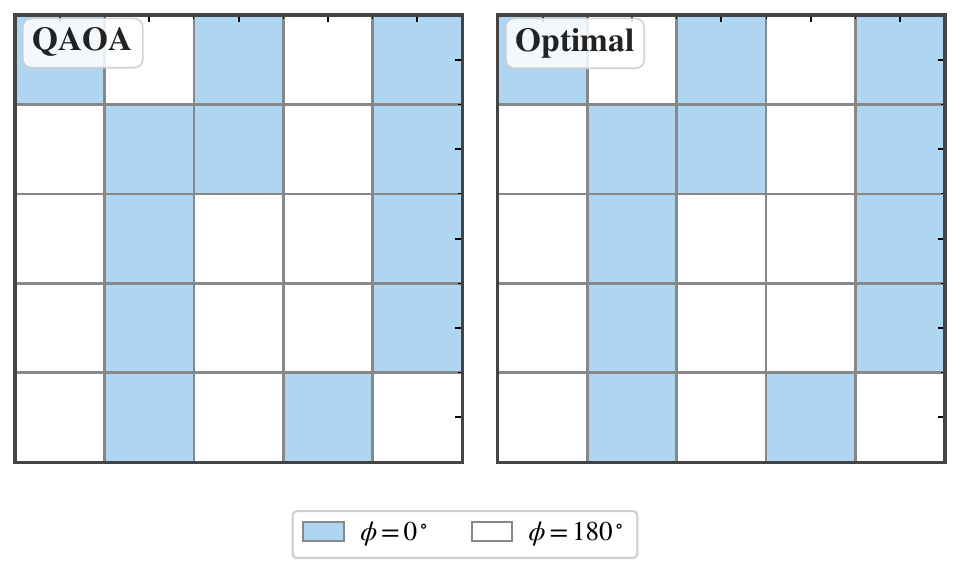}\\[0pt]
        \makebox[0.5\linewidth][c]{$\epsilon = 1.49^\circ$}%
        \makebox[0.5\linewidth][c]{$\epsilon = 1.49^\circ$}\\[4pt]
        (d) Model 4
    \end{minipage}
    
    \vspace{-0.5em}
    \caption{Results for $(\theta^{i}, \phi^{i}) = (60^\circ, 30^\circ)$ and $(\theta^{s}, \phi^{s}) = (15^\circ, 100^\circ)$. ``Optimal'' means model optimality, not physical performance. QAOA matches optimal solutions for Models 2--4, while Model 1 converges to a suboptimal configuration that better approximates the underlying physics, highlighting the mismatch between encoded $J_{ij}$ objective and full electromagnetic model.
}
    \label{fig:ris_combined_models}
\end{minipage}}
    \hfill 
    \begin{minipage}[b]{0.49\textwidth}
        \centering
        \includegraphics[width=0.95\linewidth]{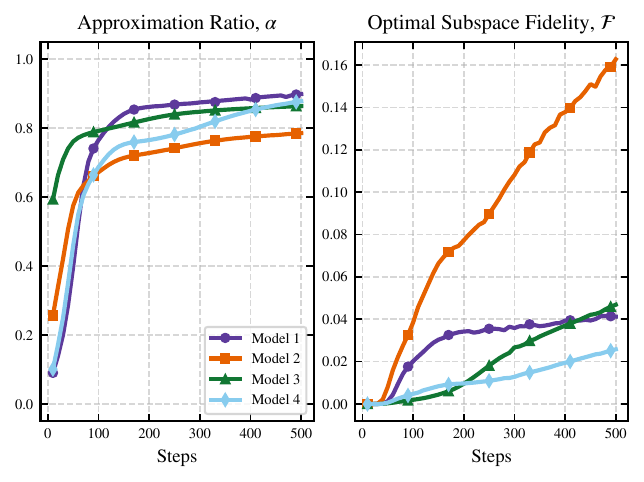}
        \\
        \vspace{-1mm}
        \makebox[0.5\linewidth][c]{(a)}%
        \makebox[0.5\linewidth][c]{(b)}

        \vspace{-3mm}
        \caption{(a) Approximation ratio (AR) computed using Eq.\eqref{eq:AR-expval} based on the expectation value of the cost Hamiltonian, demonstrating similar final AR for all four models. (b) Overlap between the quantum state and the optimal subspace, as defined in Eq.\eqref{eq:fidelity}. Unlike the AR, the overlap continues to show a clearly growing trend at the end of optimization for most models.}
        \label{fig:optimizTrends}
    \end{minipage}
    
\end{figure*}

\begin{figure*}
    \centering
    \includegraphics[width=0.975\linewidth]{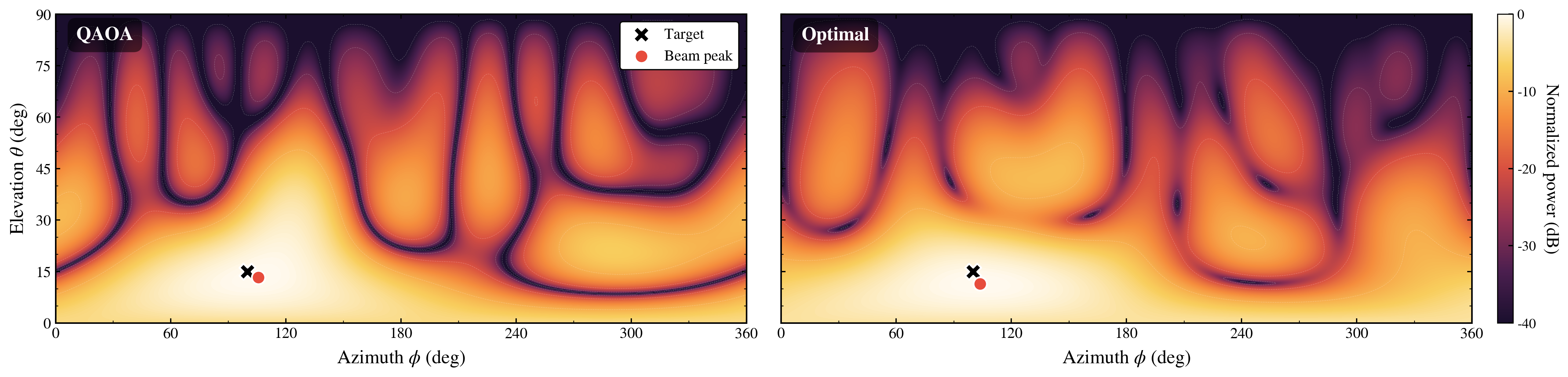}

    \vspace{-3mm}
    \caption{Radiation power $P(\theta,\phi)$ in dB, normalized to the peak scattered beam of the RIS using Model~1, comparing the QAOA-derived solution with the model-optimal solution. While QAOA does not exactly reach the optimal state (right), it still yields strong performance. Target angle is marked by a black cross, while radiation maximum is indicated in red; smaller angular separation means higher QAOA accuracy. The two solutions exhibit distinct beam patterns, including side-lobe differences.}
    \label{fig:radpatt2d}
\end{figure*}

We present the results for a $5 \times 5$-element Reconfigurable Intelligent Surface (RIS) designed to steer an incident electromagnetic wave from a specific incoming angle $(\theta^{i}, \phi^{i})$ to a target outgoing direction $(\theta^{s}, \phi^{s})$. Moving beyond idealized independent-element assumptions, the framework incorporates the models mentioned in Table~\ref{tab:qubo_models}.
The incident and target angles can be chosen arbitrarily within the physical bounds $\theta \in [0^\circ, 90^\circ]$ and $\phi \in [0^\circ, 360^\circ)$. In this study, we consider an incident direction of $(\theta^{i}, \phi^{i}) = (60.0^\circ, 30.0^\circ)$ and a target scattering direction of $(\theta^{s}, \phi^{s}) = (15.0^\circ, 100.0^\circ)$. The system operates at 30 GHz, with an element size of 5 mm (i.e., $\lambda/2$), and the QAOA optimization is performed with depth $p=6$ using the Adam optimizer.
We evaluate the performance of each model under this configuration. Each optimization is run for 500 steps, which, based on the grid size detailed in Table~\ref{tab:time}, corresponds to an execution time of approximately 6 minutes and 30 seconds per run. 

In Fig.~\ref{fig:optimizTrends}(a), we present the approximation ratio (AR) evaluated on the expectation value of the cost Hamiltonian $\braket{H_C}$, defined as
\begin{equation}
    \alpha=\frac{C_\mathrm{max}-\braket{H_C}}{C_\mathrm{max}-C_\mathrm{min}}
\label{eq:AR-expval}
\end{equation}
where $C_\mathrm{min}$ and $C_\mathrm{max}$ denote the minimum and maximum attainable costs, respectively. Additionally, in Fig.~\ref{fig:optimizTrends}(b), we evaluate the overlap of the quantum state with the optimal (ground-state) subspace. Defining the projector onto this subspace as $\Pi_{\mathrm{opt}} = \sum_{\mu=1}^{D} \ket{z^\star_\mu}\bra{z^\star_\mu}$, where $\{z^\star_\mu\}_{\mu=1}^D$ denote the $D$ degenerate optimal bitstrings (and $\ket{z^\star_\mu}$ their corresponding computational basis states), we compute
\begin{equation}
    \mathcal{F}
    = \bra{\psi(\boldsymbol{\gamma},\boldsymbol{\beta})}
    \Pi_{\mathrm{opt}}
    \ket{\psi(\boldsymbol{\gamma},\boldsymbol{\beta})}.
    \label{eq:fidelity}
\end{equation}

\begin{table}[b]
\centering
\caption{Time breakdown of a custom QAOA implementation (PyTorch, NVIDIA A100) using Model 1. Overhead includes JIT compilation and brute-force verification.}

\vspace{-2mm}
\begin{tabular}{ccc}
\hline
\textbf{Grid Size} & \textbf{Overhead (s)} & \textbf{Time per Iteration (s)} \\
\hline
3 & $14.7555 \pm 0.1764$ & $0.0081 \pm 0.0011$ \\
4 & $14.6957 \pm 0.0278$ & $0.0148 \pm 0.0005$ \\
5 & $17.0899 \pm 0.0354$ & $0.7415 \pm 0.0008$ \\
\hline
\end{tabular}
\label{tab:time}

\vspace{1mm}
{\footnotesize Averaged over 10 runs of 50 optimization steps each.}
\end{table}

While the approximation ratio (AR) measures the expected objective value over the full output distribution, the overlap $\mathcal{F}$ measures only the total probability assigned to optimal bitstrings. As a result, the most probable sampled bitstring can already be optimal while the AR still remains below one, because non-optimal bitstrings continue to carry non-negligible sampling probabilities.

In Fig.~\ref{fig:optimizTrends}(a), the AR reaches comparable values for all four models, ranging from 0.79 to 0.90. Model~2 exhibits a slight performance gap, possibly indicating a more challenging optimization landscape. The evolution of the optimal subspace overlap, shown in Fig.~\ref{fig:optimizTrends}(b), demonstrates that this sampling probability increases steadily throughout optimization for all models. Notably, while the AR nearly converges after 500 optimization steps, $\mathcal{F}$ continues to increase, suggesting longer QAOA runs can further improve the probability of sampling optimal solutions. Models with the highest $\mathcal{F}$ do not necessarily achieve the highest AR, as seen in Model~4, which has the largest optimal subspace overlap but the lowest AR. This occurs because non-optimal states with non-negligible probabilities reduce the AR while still permitting a relatively high probability of sampling an optimal solution.

From quantum circuit simulations, we observe no significant difference in the number of optimization steps required for sparse and dense QUBO instances in order to reach convergence. While the corresponding $J_{ij}$ matrices in the sparse models exhibit weaker couplings, likely leading to a smoother optimization landscape that is generally easier to navigate, Model~4, which most accurately captures the underlying physics, induces a highly nontrivial and dense optimization landscape. Notably, despite this increased complexity, Model~4 still converges rapidly to high-quality solutions, as reflected in the convergence of AR. This indicates that even physically detailed and densely coupled models can be effectively encoded and optimized within the QUBO framework.

It is particularly interesting to consider the final solution obtained for Model~1, where the most probable bitstring does not correspond to a global optimum (see Figs.~\ref{fig:ris_combined_models} and \ref{fig:radpatt2d}). Notably, this suboptimal QUBO solution still yields physically accurate results, with an angular error of merely $\epsilon$ = 2.23°, which is even slightly lower than the $\epsilon$ = 3.66° obtained for the optimal bitstring. \emph{This provides a clear example that optimality with respect to the QUBO objective does not necessarily translate directly into improved physical accuracy}, highlighting a nontrivial mismatch between the optimization landscape and the underlying physical objective.

Lastly, we briefly comment on the mapping of the different models to hardware. To quantify the \textit{quantum suitability} in Table~\ref{tab:qubo_models}, we estimate the gate overhead required to implement each model on a Noisy Intermediate-Scale Quantum (NISQ) processor. For a $5 \times 5$ RIS grid ($N_Q = 25$ qubits), the dense Hamiltonians of Models~1 and~4 require all-to-all logical qubit connectivity, corresponding to $N_Q(N_Q-1)/2 = 300$ pairwise interactions. In contrast, Models~2 and~3 impose spatial cutoffs, reducing the interaction graph to nearest-neighbor couplings (approximately $40$ interactions for a $5 \times 5$ grid). Such sparse interaction structures are naturally compatible with common hardware connectivity graphs, thereby substantially reducing the need for qubit routing and the associated gate overhead. Consequently, while Model~4 most faithfully captures the underlying physical coupling, Models~2 and~3 offer a more favorable trade-off for near-term implementations.

\section{Limitations and Future Work}
Although this work successfully connects quantum optimization with electromagnetic coupling, several constraints remain, including hardware limitations, algorithmic challenges, and physical modeling assumptions.

A first limitation lies in the physical constraints of current Noisy Intermediate-Scale Quantum (NISQ) devices. Accurately mapping dense electromagnetic interactions yields cost Hamiltonians with all-to-all connectivity, bringing significant qubit routing overhead on near-term hardware. This leads to deeper circuits and increases required QAOA layers ($p$), making the algorithm highly susceptible to noise and decoherence, thus degrading solution quality.

On the algorithmic side, QAOA does not guarantee convergence to the global optimum, although in this case study it reliably reaches optimal configurations. Its performance depends strongly on the choice of classical optimizer (e.g., L-BFGS-B, Adam, COBYLA, SPSA) and initialization. As the RIS grid scales, the optimization landscape becomes more challenging, with barren plateaus emerging. In these regions, gradients vanish exponentially, making training difficult and often leading to convergence at suboptimal local minima. Moreover, we assume access to exact gradients; on real hardware, gradients must be estimated via sampling-based methods such as the parameter-shift rule~\cite{crooks_gradients_2019}. This introduces further complications, as finite sampling and hardware noise can degrade gradient quality.

From an electromagnetic standpoint, while the spherical-wave mutual coupling model used to benchmark accuracy improves upon isolated-array approximations, it remains an abstraction. A fully accurate evaluation should incorporate geometric constraints and edge effects, relying on full-wave electromagnetic simulations or scattering parameter ($S$-parameter) matrices \cite{zheng_mutual_2024}. However, such modeling strategies are computationally expensive, motivating the development of surrogate models to efficiently approximate system behavior. Moreover, $J_{ij}$ should account for sidelobe suppression in $P(\theta, \phi)$, as sidelobes are undesirable due to power losses and reduced beamforming efficiency. Lastly, scaling RIS systems to sizes relevant for 6G applications poses an additional challenge. Even moderate array sizes lead to exponentially large configuration spaces, making brute-force verification unfeasible. To address this, future work may explore alternative methods to efficiently represent and simulate large-scale systems, enabling the study of more realistic RIS configurations beyond current quantum hardware limitations.

\section{Conclusion}

In this paper, we introduced a QUBO modeling framework for RIS optimization with QAOA that incorporates far-field interference and mutual-coupling effects into the Ising Hamiltonian. This yields models that are more physically informed than standard toy formulations, while still remaining simplified electromagnetic abstractions. Evaluating four distinct interaction ($J_{ij}$) models yielded promising results, demonstrating the efficiency of quantum optimization in this problem space. The Adam-driven hybrid quantum–classical framework proved effective for tested configurations, such as steering an incident wave from $(\theta^i, \phi^i)$ = (60°, 30°) to a target of $(\theta^s, \phi^s)$ = (15°, 100°). For the physics-informed models (Models 2, 3, and 4), QAOA rapidly converged to the optimal phase configuration, with the most probable measured bitstring matching the global minimum identified via exhaustive search.

A key finding of this work is the optimizer's robustness when faced with highly realistic electromagnetic constraints. Although the fully coupled physical model (Model 4) yields a highly complex and dense optimization landscape, the QAOA routine still achieved an approximation ratio comparable to or better than those obtained for simpler models. This encouraging outcome indicates that formulating detailed, dense physical models within a QUBO framework does not necessarily deteriorate QAOA optimization performance. 

While our results highlight a trade-off between denser electromagnetic modeling and the sparse connectivity constraints of current NISQ devices, they should be interpreted within the simplified setting studied here. Within this scope, QAOA was able to optimize coupled QUBO instances and recover beamforming configurations performing well under the validation model. These findings reveal that introducing more physics-informed structure into quantum optimization formulations is promising, though substantial steps remain before such approaches become realistic RIS design tools.

\begin{acks}
This work has received funding from the Swedish Research Council’s Research Environment grant (SEE-6GIA 2024-06482). We also acknowledge the Wallenberg Centre for Quantum Technology (WACQT) for access to their platforms.
\end{acks}

\bibliographystyle{ACM-Reference-Format}
\bibliography{acmart}

@article{colella_quantum_2024,
    title = {Quantum {Optimization} of {Reconfigurable} {Intelligent} {Surfaces} for {Mitigating} {Multipath} {Fading} in {Wireless} {Networks}},
    volume = {9},
    issn = {2379-8793},
    url = {https://ieeexplore.ieee.org/document/10747251/},
    doi = {10.1109/JMMCT.2024.3494037},
    urldate = {2026-03-04},
    journal = {IEEE Journal on Multiscale and Multiphysics Computational Techniques},
    author = {Colella, Emanuel and Bastianelli, Luca and Primiani, Valter Mariani and Peng, Zhen and Moglie, Franco and Gradoni, Gabriele},
    year = {2024},
}

@article{batool_quantum-enhanced_2026,
    title = {Quantum-{Enhanced} {Massive} {MIMO} {Beamforming} for {6G} {IoT} {Networks}: {A} {QAOA}-{Based} {Optimization} {Framework}},
    volume = {7},
    issn = {2644-125X},
    shorttitle = {Quantum-{Enhanced} {Massive} {MIMO} {Beamforming} for {6G} {IoT} {Networks}},
    url = {https://ieeexplore.ieee.org/document/11303130},
    doi = {10.1109/OJCOMS.2025.3645207},
    urldate = {2026-02-26},
    journal = {IEEE Open Journal of the Communications Society},
    author = {Batool, Iqra and Fouda, Mostafa M. and Ismail, Muhammad and Ibrahem, Mohamed I. and Md Fadlullah, Zubair and Kato, Nei},
    year = {2026},
}

@inproceedings{choi_tutorial_2019,
    title = {A {Tutorial} on {Quantum} {Approximate} {Optimization} {Algorithm} ({QAOA}): {Fundamentals} and {Applications}},
    issn = {2162-1233},
    shorttitle = {A {Tutorial} on {Quantum} {Approximate} {Optimization} {Algorithm} ({QAOA})},
    url = {https://ieeexplore.ieee.org/document/8939749/},
    doi = {10.1109/ICTC46691.2019.8939749},
    urldate = {2026-02-26},
    booktitle = {2019 {International} {Conference} on {Information} and {Communication} {Technology} {Convergence} ({ICTC})},
    author = {Choi, Jaeho and Kim, Joongheon},
    year = {2019},
    note = {ISSN: 2162-1233},
}

@article{ross_engineering_2022,
    title = {Engineering {Reflective} {Metasurfaces} {With} {Ising} {Hamiltonian} and {Quantum} {Annealing}},
    volume = {70},
    issn = {1558-2221},
    url = {https://ieeexplore.ieee.org/document/9665281},
    doi = {10.1109/TAP.2021.3137424},
    number = {4},
    urldate = {2026-02-25},
    journal = {IEEE Transactions on Antennas and Propagation},
    author = {Ross, Charles and Gradoni, Gabriele and Lim, Qi Jian and Peng, Zhen},
    year = {2022},
}

@article{lee_quantum_2025,
    title = {Quantum {Annealing} for {Electromagnetic} {Engineers}—{Part} {I}: {A} computational method to solve various types of optimization problems.},
    volume = {67},
    copyright = {https://ieeexplore.ieee.org/Xplorehelp/downloads/license-information/IEEE.html},
    issn = {1045-9243, 1558-4143},
    shorttitle = {Quantum {Annealing} for {Electromagnetic} {Engineers}—{Part} {I}},
    url = {https://ieeexplore.ieee.org/document/10798976/},
    doi = {10.1109/MAP.2024.3498695},
    language = {en},
    number = {6},
    urldate = {2026-02-25},
    journal = {IEEE Antennas and Propagation Magazine},
    author = {Lee, Sangbin and Lim, Qi Jian and Ross, Charles and Lee, Eungkyu and Han, Soyul and Kim, Youngmin and Peng, Zhen and Kim, Sanghoek},
    year = {2025},
}

@article{lee_quantum_2026,
    title = {Quantum {Annealing} for {Electromagnetic} {Engineers}—{Part} {II}: {Examples} of electromagnetic problems solved by quantum annealing},
    volume = {68},
    issn = {1558-4143},
    shorttitle = {Quantum {Annealing} for {Electromagnetic} {Engineers}—{Part} {II}},
    url = {https://ieeexplore.ieee.org/document/10873815/},
    doi = {10.1109/MAP.2025.3530408},
    number = {1},
    urldate = {2026-02-24},
    journal = {IEEE Antennas and Propagation Magazine},
    author = {Lee, Sangbin and Lim, Qi Jian and Ross, Charles and Lee, Eungkyu and Han, Soyul and Kim, Youngmin and Peng, Zhen and Kim, Sanghoek},
    year = {2026},
}

@misc{farhi_quantum_2014,
    title = {A {Quantum} {Approximate} {Optimization} {Algorithm}},
    url = {http://arxiv.org/abs/1411.4028},
    doi = {10.48550/arXiv.1411.4028},
    urldate = {2025-01-10},
    publisher = {arXiv},
    author = {Farhi, Edward and Goldstone, Jeffrey and Gutmann, Sam},
    year = {2014},
    note = {arXiv:1411.4028},
}

@article{liu_reconfigurable_2021,
    title = {Reconfigurable {Intelligent} {Surfaces}: {Principles} and {Opportunities}},
    volume = {23},
    issn = {1553-877X},
    shorttitle = {Reconfigurable {Intelligent} {Surfaces}},
    url = {https://ieeexplore.ieee.org/document/9424177},
    doi = {10.1109/COMST.2021.3077737},
    number = {3},
    urldate = {2026-04-15},
    journal = {IEEE Communications Surveys \& Tutorials},
    author = {Liu, Yuanwei and Liu, Xiao and Mu, Xidong and Hou, Tianwei and Xu, Jiaqi and Di Renzo, Marco and Al-Dhahir, Naofal},
    year = {2021},
}

@article{rana_review_2023,
    title = {Review {Paper} on {Hardware} of {Reconfigurable} {Intelligent} {Surfaces}},
    volume = {11},
    issn = {2169-3536},
    url = {https://ieeexplore.ieee.org/abstract/document/10080950},
    doi = {10.1109/ACCESS.2023.3261547},
    urldate = {2026-04-15},
    journal = {IEEE Access},
    author = {Rana, Biswarup and Cho, Sung-Sil and Hong, Ic-Pyo},
    year = {2023},
}

@article{elmossallamy_reconfigurable_2020,
    title = {Reconfigurable {Intelligent} {Surfaces} for {Wireless} {Communications}: {Principles}, {Challenges}, and {Opportunities}},
    volume = {6},
    issn = {2332-7731},
    shorttitle = {Reconfigurable {Intelligent} {Surfaces} for {Wireless} {Communications}},
    url = {https://ieeexplore.ieee.org/abstract/document/9086766},
    doi = {10.1109/TCCN.2020.2992604},
    number = {3},
    urldate = {2026-04-15},
    journal = {IEEE Transactions on Cognitive Communications and Networking},
    author = {ElMossallamy, Mohamed A. and Zhang, Hongliang and Song, Lingyang and Seddik, Karim G. and Han, Zhu and Li, Geoffrey Ye},
    year = {2020},
}

@book{punnen_quadratic_2022,
    address = {Cham},
    title = {The {Quadratic} {Unconstrained} {Binary} {Optimization} {Problem}: {Theory}, {Algorithms}, and {Applications}},
    copyright = {https://www.springer.com/tdm},
    isbn = {978-3-031-04519-6 978-3-031-04520-2},
    shorttitle = {The {Quadratic} {Unconstrained} {Binary} {Optimization} {Problem}},
    url = {https://link.springer.com/10.1007/978-3-031-04520-2},
    doi = {10.1007/978-3-031-04520-2},
    language = {en},
    urldate = {2025-01-10},
    publisher = {Springer International Publishing},
    editor = {Punnen, Abraham P.},
    year = {2022},
}

@book{balanis_balanis_2023,
    edition = {1},
    title = {Balanis' {Advanced} {Engineering} {Electromagnetics}},
    isbn = {978-1-394-18001-1 978-1-394-18004-2},
    url = {https://onlinelibrary.wiley.com/doi/book/10.1002/9781394180042},
    doi = {10.1002/9781394180042},
    language = {en},
    urldate = {2025-10-31},
    publisher = {Wiley},
    author = {Balanis, Constantine A.},
    year = {2023},
}

@misc{asadi_hybrid_2024,
    title = {Hybrid quantum programming with {PennyLane} {Lightning} on {HPC} platforms},
    url = {http://arxiv.org/abs/2403.02512},
    doi = {10.48550/arXiv.2403.02512},
    urldate = {2026-04-21},
    publisher = {arXiv},
    author = {Asadi, Ali and Dusko, Amintor and Park, Chae-Yeun and Michaud-Rioux, Vincent and Schoch, Isidor and Shu, Shuli and Vincent, Trevor and O'Riordan, Lee James},
    year = {2024},
    note = {arXiv:2403.02512},
}

@misc{kingma_adam_2017,
    title = {Adam: {A} {Method} for {Stochastic} {Optimization}},
    shorttitle = {Adam},
    url = {http://arxiv.org/abs/1412.6980},
    doi = {10.48550/arXiv.1412.6980},
    urldate = {2026-04-21},
    publisher = {arXiv},
    author = {Kingma, Diederik P. and Ba, Jimmy},
    year = {2017},
    note = {arXiv:1412.6980},
}

@misc{crooks_gradients_2019,
    title = {Gradients of parameterized quantum gates using the parameter-shift rule and gate decomposition},
    url = {http://arxiv.org/abs/1905.13311},
    doi = {10.48550/arXiv.1905.13311},
    urldate = {2025-01-27},
    publisher = {arXiv},
    author = {Crooks, Gavin E.},
    year = {2019},
    note = {arXiv:1905.13311},
}

@article{zheng_mutual_2024,
    title = {Mutual {Coupling} in {RIS}-{Aided} {Communication}: {Model} {Training} and {Experimental} {Validation}},
    volume = {23},
    copyright = {https://ieeexplore.ieee.org/Xplorehelp/downloads/license-information/IEEE.html},
    issn = {1536-1276, 1558-2248},
    shorttitle = {Mutual {Coupling} in {RIS}-{Aided} {Communication}},
    url = {https://ieeexplore.ieee.org/document/10669140/},
    doi = {10.1109/TWC.2024.3451548},
    language = {en},
    number = {11},
    urldate = {2026-04-22},
    journal = {IEEE Transactions on Wireless Communications},
    author = {Zheng, Pinjun and Wang, Ruiqi and Shamim, Atif and Al-Naffouri, Tareq Y.},
    year = {2024},
}

\end{document}